\documentclass[12pt]{article}

\renewcommand{\baselinestretch}{1.7}

\usepackage[dvips]{color}

\usepackage{graphicx}

\usepackage{amsmath}
\usepackage{amssymb}
\usepackage{epsf}
\usepackage{latexsym}

\newcommand{\be}{\begin{equation}}
\newcommand{\ee}{\end{equation}}
\newcommand{\bea}{\begin{eqnarray}}
\newcommand{\eea}{\end{eqnarray}}

\begin{document}
\begin{titlepage}
\null
\begin{flushright}
%hep-th/YYMMXXX
%HIP-2007-62/TH
\end{flushright}

\vskip 1.2cm

\begin{center}
%{\Huge \bf Preliminary version 1, 17.11.2009}

  {\Large \bf On Finite Noncommutativity in Quantum Field Theory}

\vskip 2.0cm
\normalsize

  {\large Miklos L\aa ngvik\footnote{miklos.langvik@helsinki.fi} and Ali Zahabi\footnote{seyedali.zahabi@helsinki.fi}}

\vskip 0.5cm

  {\large \it Department of Physics, University of Helsinki, \\
              P.O. Box 64, FIN-00014 Helsinki, Finland}

\vskip 1cm

{\bf Abstract}

\end{center}

\renewcommand{\baselinestretch}{1.5}\selectfont

We consider various modifications of the Weyl-Moyal star-product, in order to obtain a finite range of nonlocality. The basic requirements are to preserve the
commutation relations of the coordinates as well as the associativity of the new product.
We show that a modification of the differential representation of the Weyl-Moyal star-product by an exponential function of derivatives will not lead to a
finite range of nonlocality. We also modify the integral kernel of the star-product introducing a Gaussian damping, but find a nonassociative
product which remains infinitely nonlocal. We are therefore led to propose that the Weyl-Moyal product should be modified by a cutoff like function,
in order to remove the infinite nonlocality of the product. We provide such a product, but it appears that one has to abandon the possibility of analytic calculation
with the new product.

\end{titlepage}

\newpage

\section{Introduction \label{intro}}

In a noncommutative field theory the space-time coordinates $x_{\mu}$ are promoted to the status of operators $\hat{x}_{\mu}$
which are characterized by the commutation relation
\be
[\hat{x}_{\mu},  \hat{x}_{\nu}] = i\theta_{\mu\nu}. \label{ncfi}
\ee
The commutation relation \eqref{ncfi} is motivated by the works \cite{DopFreRob} where $\theta_{\mu\nu}$ is taken as a tensor and \cite{SeiWitt}, where $\theta_{\mu\nu}$ is taken as a constant
antisymmetric matrix. In \cite{DopFreRob}, it was shown that
the commutation relation \eqref{ncfi} can follow from a high energy gedanken experiment where, as the energy is raised, the formation
of black holes prevents smaller distances than the diameter of the black hole to be measured. Then space-time can be characterized
by uncertainty relations among coordinate operators, which can straightforwardly be interpreted as noncommutativity of
space-time coordinates. In \cite{SeiWitt}, the commutation relation \eqref{ncfi} follows as a low-energy limit of open string theory in a background
$B$-field.

Although noncommutative field theories have a solid motivation in the works \cite{DopFreRob} and \cite{SeiWitt}, they exhibit certain
problematic properties such as the UV/IR mixing effect \cite{MinRaaSei} and the apparent violation of Lorentz invariance by the commutator \eqref{ncfi}. The problem of Lorentz noninvariance
does nevertheless not result in a need to change the representations of the Poincar\'e group  \cite{ChaKulNisTur} and consequently, the classification of particles
according to their spin and mass in commutative quantum field theory can be taken over to noncommutative quantum field theory without change. This justification is very important
for the usage of Lorentz invariant quantities and representations of the Poincar\'e symmetry within noncommutative field theory. Therefore one might
say that the problem of Lorentz noninvariance, within noncommutative field theory, is presently a smaller problem than the one of UV/IR mixing.
Indeed, the UV/IR mixing effect has yet to be included in a theory, i.e. so that it does not spoil renormalizability, or alternatively by some unknown mechanism,
completely removed from the theory. One can generally argue that the UV/IR mixing effect is a result of the infinite nonlocality of the commutator \eqref{ncfi}
and that restricting the range of it, should result in a theory without UV/IR mixing.

Work along these lines has been done in \cite{Bahns:2006cp} where it was attempted to come to terms with the infinite nonlocality of noncommutative
models of space-time by introducing a support for the noncommutativity parameter $\theta$ inside a specific range. This, together with an appropriately chosen
deformation of the states of the theory, results in a finite range for
microcausality as it reduces to the microcausality of commutative quantum field theory outside the support of $\theta$. However, this
approach makes it difficult to construct an interaction that would remain nonlocal on a finite range and moreover, the definition of observables that
respect this type of microcausality becomes nontrivial. In addition, the choice of the deformation of the states is highly nonunique.The present work differs from the approach in \cite{Bahns:2006cp}, in that no
support for $\theta$ is introduced, and were we able to construct a product under our requirements, there would be no problem
of introducing interactions or observables into the theory.

The work is organized as follows. We begin by giving the representations of the star-product most useful for this work in section \ref{repstar}.
We then move on to modify the star product with a Gaussian damping term and discuss the possibility of modifying the
differential representation of the star-product. This analysis leads us to propose a modification of the integral-kernel of the
Weyl-Moyal star-product by a Heaviside stepfunction cutoff in section \ref{modstar}. In section \ref{conclrem} we
make our concluding remarks.

\section{Representations of the star-product \label{repstar}}

In the usual approach to noncommutative quantum field theory we assume the commutator
\be
[\hat{x}_{\mu},  \hat{x}_{\nu}] = i\theta_{\mu\nu}, \label{comx}
\ee
where $\theta_{\mu\nu}$ is taken to be a constant and antisymmetric matrix. However, because of the loss of
unitarity \cite{Gomis:2000zz} and causality \cite{Seiberg:2000gc, ChaNisTur}, in theories where time and space do not commute,
one often considers theories with only space-space noncommutativity. This will also be the approach in this work. Additionally, since $\theta_{\mu\nu}$ can always be transformed into a frame where
only four distinct components of the antisymmetric matrix survive and only two of them are independent, we take
$\theta_{\mu\nu}$ to be given by
\begin{eqnarray}
\theta_{23}=\theta\neq0,~~ \theta_{12}= \theta_{13}= \theta_{0i}=0,
\label{plane}
\end{eqnarray}
so that the symmetry of space-time is $O(1,1)\times SO(2)$.
The standard way to realize the commutator \eqref{comx} is via the Weyl-Moyal star-product
\begin{eqnarray}
(fg)(\hat{x})\longmapsto(f*_W g)(x) =
  \textrm{exp}\left[\frac{i}{2}\theta^{\mu\nu}
    {\partial\over \partial x^\mu}{\partial\over \partial y^\nu}\right]
f(x)g(y)
 \bigg| _{x=y}. \label{star}
\end{eqnarray}
That is, products between functions of $x$ are now taken with respect to the star-product \eqref{star}.
Another way is to use the coherent-state basis.
This corresponds to optimal localization in the noncommutative plane
and leads to the Wick-Voros  star-product,
\begin{eqnarray}
(f*_V g)(x)= e^{\frac{i\theta}{2}\partial_x\wedge\partial_y+\frac{\theta}{2}(\partial_x\partial_y)}f(x) g(y)\bigg|_{y=x},
\label{WV}
\end{eqnarray}
where $\partial_x\wedge\partial_y=\frac{\partial}{\partial x_2}\frac{\partial}{\partial y_3}-\frac{\partial}{\partial x_3}\frac{\partial}{\partial y_2}$ and  $\partial_x\partial_y=\frac{\partial}{\partial x_2}\frac{\partial}{\partial y_2}+\frac{\partial}{\partial x_3}\frac{\partial}{\partial y_3}$.

These star-products give isomorphic representations of the algebra of noncommutative fields.
In the extensively studied $*_W$ -case one encounters nonlocality-related problems such as UV/IR-mixing and acausality, arising from
the fact that the product is infinitely nonlocal. This is clearly seen from the integral representation of the product:
\begin{eqnarray}
f(x)*_W g(x)=\int d^D z\;d^D y \; \frac{1}{\pi^D \det \theta} \exp[2i(x\theta^{-1}y+y\theta^{-1}z+z\theta^{-1}x)] f(y) g(z),
\label{int1}
\end{eqnarray}
which receives contributions from $f(y)$ and $g(z)$ for all values of $y$ and $z$.
Thus, for example the matrix elements of the commutator
\begin{eqnarray}
[:\phi(x)*\phi(x):,:\phi(y)*\phi(y):],
\label{microc}
\end{eqnarray}
vanish in general only for $((x_0-y_0)^2-(x_1-y_1)^2)<0$, corresponding to the light-wedge causality condition \cite{ChaNisTur, AlvarezGaume:2001ka}.
% (it is nonvanishing everywhere if time is allowed to be noncommutative) [].
%is in general nonvanishing for all $|x-y|$, leading to complete loss of causality
%(or at least modified causality condition in the case of $\theta^{0i}=0$).

In the case of $*_V$ these problems still persist. It is notable that the different
form of the star-product leads to damping factors in the Green functions. For example the
propagator of a free scalar field has a damping factor that makes the Green function finite:
\begin{eqnarray}
G^{(0)}(x,y)=\frac{1}{(2 \pi )^3} \int d^3 k e^{ik(x-y)}e^{-\frac{\theta}{2} k^2}
\label{}
\end{eqnarray}
However, in perturbation theory, the damping factors in the propagators are cancelled by opposite factors from the vertices,
so that UV divergences and UV/IR mixing still appear \cite{ChaDemPre}.
%As it was mentioned in  \cite{ChaDemPre}, one way to obtain a UV-regular theory is to use Weyl symbols in the
%interaction terms and normal symbols in the free action. However, there is no principle for preferring this ordering,
%except a pragmatic point of view.

\section{Modifying the star product \label{modstar}}

The right hand side of the commutator (\ref{comx}) has to be a constant in order to preserve translational invariance. That is why we will adopt the view that we cannot modify the
r.h.s. to obtain noncommutativity of a finite range, since any $x$-dependent function would break translational invariance. Nevertheless, we may note that similar considerations of modifying
the star-product to obtain a new associative product when $\theta$ is a function of the position $x$, have been made in \cite{Gayral:2005ih}.

\subsection{Modified star-product \label{modsec}}

Instead of modifying the commutator itself, let us now consider modifying the star-product.
We will generally require that our modified star-product respects the commutator \eqref{comx}
and remains associative. We will also assume, for the sake of simplicity, that the noncommutativity is restricted to a plane as in \eqref{plane}.

Let us begin by introducing some damping into the integrand of \eqref{int1}. For example an
exponential damping, $\exp[-l^{-2}(x-y)^2-l^{-2}(x-z)^2]$, where $l$ has a finite value and represents the scale for
the reach of the noncommutativity.

\subsubsection{Gaussian damping \label{gauda}}

In the case of a Gaussian damping of the star-product,
we can, using (\ref{int1}), define the modified star-product as
\begin{eqnarray}
f(x)*'g(x)&:=\int d^2 z\;d^2 y\;\frac{1}{\pi^2 \det \theta} \exp[\frac{2i}{\theta}(x\wedge y+y\wedge z+z\wedge x)] \nonumber\\
&\exp[-\frac{1}{\theta}\left((x-y)^2+(x-z)^2\right)] \; f(y) \; g(z),
\label{integral}
\end{eqnarray}
where we have denoted $(x_2,x_3)$ and $(y_2, y_3)$ by $x$ and $y$ respectively and taken $l^2 = \theta$.
Now we should check our requirements stated at the beginning of section \ref{modsec}, e.g. to what extent it realizes the
commutator \eqref{comx} and whether it gives an associative product.

For this kind of a modification of the star-product we have the following results:

1. The product of two plane waves is up to a multiplicative constant
\begin{eqnarray}
e^{ix\cdot k}*'e^{ix\cdot q}=e^{ix\cdot (k+q)} e^{\frac{-i\theta}{2}k\wedge q} e^{\frac{-\theta}{4}(k^2+q^2)}.
\label{exp}
\end{eqnarray}
This is to be contrasted with the standard star-product case,
\begin{eqnarray}
e^{ix\cdot k} *_W e^{ix\cdot q}=e^{ix\cdot (k+q)}e^{-\frac{i}{2}\theta k\wedge q},
\label{}
\end{eqnarray}
and the Wick-Voros case:
\begin{eqnarray}
e^{ix\cdot k} *_V e^{ix\cdot q} = e^{ix\cdot (k+q)} e^{-\frac{i}{2}\theta k\wedge q} e^{-\frac{\theta}{2} k\cdot q}.
\label{}
\end{eqnarray}

2. By considering the product of three plane waves
\begin{eqnarray}
e^{ix\cdot p}*'(e^{ix\cdot k}*'e^{ixq})&=& e^{ix\cdot (p+k+q)} e^{\frac{-i\theta}{2} p\wedge (k+q)} e^{\frac{-\theta}{4}(p^2+(k+q)^2)} \left(e^{\frac{-i\theta}{2}k\wedge q} e^{\frac{-\theta}{4}(k^2+q^2)}\right) \nonumber \\
&\not=& (e^{ix\cdot p}*'e^{ix\cdot k})*'e^{ix\cdot q},
\label{}
\end{eqnarray}
we see that the product is nonassociative.

3. It realizes the commutator $[x_2, x_3]_{*'}=i\theta$.

Equation (\ref{exp}) suggests that the product can be written as an exponential of derivatives as
\begin{eqnarray}
(f*_mg)(x)= e^{\frac{i\theta}{2}\partial_x\wedge\partial_y+\frac{\theta}{4}(\partial_x^2+\partial_y^2)}f(x)g(y)\bigg|_{y=x},
\label{derivative}
\end{eqnarray}
where $*_m$ stands for a modified star-product. This resembles the Wick-Voros product (\ref{WV})
However, the Wick-Voros product is associative and it seems that the non-associativity of (\ref{derivative}) arises from the second order derivatives with respect
to the same variable in the exponent of (\ref{derivative}). We will show that this is indeed the case in section \ref{expmod}.

In order to see the effect of the Gaussian damping to microcausality, we calculate the Equal Time Commutation Relation (ETCR)
\begin{eqnarray}
<0|[:\phi(x)*'\phi(x):,:\phi(y)*'\phi(y):]|p,p'>.
\label{}
\end{eqnarray}

In the case of $\theta_{0i}=0$, the ETCR vanishes as in the case of the usual star-product. So let us suppose
a 1+1 dimensional space-time with $\theta_{01}=\theta>0$, and define
the $*'$ product similarly to the $\theta_{0i}=0$ case, by:
\begin{eqnarray}
f(x)*'g(x)&:=\int d^2 z\;d^2 y\;\frac{1}{\pi^2 \det \theta} \exp[\frac{2i}{\theta}(x\wedge y+y\wedge z+z\wedge x)] \nonumber\\
&\exp[-\frac{1}{\theta}\left((x-y)^2+(x-z)^2\right)] \; f(y) \; g(z),
\label{integral2}
\end{eqnarray}
with this definition the ETCR for a Euclidean signature in the damping factors $(x-y)^2 :=(x^0-y^0)^2 + (x^1-y^1)^2$ and
$(x-z)^2 :=(x^0-z^0)^2 + (x^1-z^1)^2$ becomes, up to a constant factor,
\begin{eqnarray}
(e^{-ip'y-ipx}+e^{-ip'x-ipy}) \int \;d^2 k \;e^{-\frac{\theta}{4}(2k^2+p^2+p'^2)}\;\cos(\frac{\theta}{2}p\wedge k)\cos(\frac{\theta}{2}p'\wedge k)\sin(k(y-x)),
\label{etcrnc}
\end{eqnarray}
where the sign in the Gaussian factor changes to + for a Minkowskian signature in the damping factor.
For the Minkowskian signature the product may however, not be well defined, as the Gaussian exponential blows up for $y^2\longrightarrow -\infty$ or $z^2\longrightarrow -\infty$ in \eqref{integral2} and
we therefore consider only the Euclidean case here.

From the form \eqref{etcrnc} we see that the ETCR certainly does not vanish even for large $(y - x)^2$. This was to be expected, because
of the infinite tails of the Gaussian distribution. That is, although small, the Gaussian distribution has some value even at very large $x$ and vanishes strictly first at infinity. This suggests that
a modification of the star-product, that would produce a strictly vanishing ETCR for $(x - y)^2<-l^2$, should involve some sort of cutoff. For example a step-function
type of cutoff. We will explore this possibility in section \ref{step}. An apparent drawback of the star-product modified by a Gaussian damping is the loss of associativity, due to
which we cannot use it straightforwardly to define a sensible field theory in noncommutative space-time. It would 
nonetheless be interesting to know if the nonassociativity of this particular product could have been expected from a geometrical
interpretation of the associativity of star-products, as in \cite{zach}.

\subsubsection{An exponential modification of the star-product  \label{expmod}}

If we define some modification of the Weyl-Moyal product as
\begin{eqnarray}
(f*_mg)(x)=e^{\frac{i\theta}{2}\partial_x\wedge\partial_y+F(\partial_x,\partial_y)}f(x) g(y)\bigg|_{y=x} ,
\label{modprod}
\end{eqnarray}
where $F(\partial_x, \partial_y)$ is an arbitrary function of the differentials $\partial_x$ and $\partial_y$ that contains $\theta$ to first order,
then the requirement of associativity can be reduced to the equation
\be
\Big[F(\partial_y, \partial_z)f(y)g(z)h(z) + f(x)F(\partial_y, \partial_z)g(y)h(z) = F(\partial_y, \partial_z)f(y)g(z)h(x)+F(\partial_y, \partial_z)f(y)g(y)h(z)\Big]\bigg|_{y=z=x},
\ee
which is the equation of associativity for the first order terms in the expansion of the exponent of the product $*_m$ in \eqref{modprod}. We can easily see from here that
only in the case that the derivative operator $F(\partial_y,\partial_z)$ is linear in both $\partial_y$ and $\partial_z$, but does
not contain higher order derivatives of the same variable, the equation is satisfied. This clearly shows that any attempt to modify the differential representation of the Weyl-Moyal star-product by
an exponential function of derivatives, must be of the form
\be
(f*_mg)(x)=e^{\frac{i\theta}{2}\partial_x\wedge\partial_y+ a^{ij}\partial^i_x\partial^j_y + b^i\partial^i_x + c^i\partial^i_y}f(x) g(y)\bigg|_{y=x},
\label{fmg}
\ee
to remain associative. Here, $a^{ij}, b^i$ and $c^i$ are constants, independent of $y, z$ or $x$. If we also require that the product produces the
commutator \eqref{comx}, we are led to the requirement
\be
a^{ij}  =  a^{ji} \label{symm}, \hspace{10pt} b^i  =   c^i.
\ee
If we calculate the product of two plane waves using \eqref{fmg} with the requirement \eqref{symm}, we obtain
\be
e^{ik\cdot x}*_m e^{iq\cdot x} = e^{ik\cdot(x + b) + iq\cdot(x + b)}e^{-\frac{i}{2}\theta k\wedge q}e^{-a^{ij}k_iq_j}.
\ee
From this it can be seen that the terms with $b^i \neq 0$, only correspond to a translation of the coordinates $x$ by a constant vector $b^i$, and therefore
we will omit them from now on. The $a^{ij}$-part of the
product can on the other hand be recognized as the Wick-Voros product in the particular case: $a^{ij} = \frac{\theta}{2}\delta^{ij}$. Therefore, 
we should check whether this modified star-product can change the properties of microcausality in noncommutative quantum field theory, should
we be able to choose $a^{ij}$ appropriately.

The ETCR for the star-product \eqref{fmg} with $b^i = c^i = 0$ and $a^{ij} = a^{ji}$ yields
\be
(e^{-ip'y-ipx}+e^{-ip'x-ipy}) \int \;d^2 k \;e^{-a^{ij}(p'_i + p_i)k_j}\;\cos(\frac{\theta}{2}p\wedge k)\cos(\frac{\theta}{2}p'\wedge k)\sin(k(y-x)).
\label{symetcr}
\ee
Clearly, there is no way of making the expression \eqref{symetcr} vanish for any $(y-x)^2$ no matter how we choose $a^{ij}$. This was to be expected,
knowing that the Wick-Voros product does not change the locality properties of field theory and it is a special case
of \eqref{fmg} with $a^{ij} = \frac{\theta}{2}\delta^{ij}$ and $b^i = c^i = 0$. At best, an appropriately chosen $a^{ij}$ could introduce some small suppression
into the ETCR, but since it would not solve the UV/IR problem \cite{Galluccio:2009ss}, we will not investigate the properties of \eqref{fmg} any further.
What is appealing with the product \eqref{fmg} under the requirement \eqref{symm}, is that it is associative and satisfies the commutator \eqref{comx}.

\subsubsection{Step function \label{step}}
If we want to obtain strict finite locality at some range, it seems that we must introduce a cutoff-like function
into the integral representation of the star-product \eqref{int1}, as discussed in section \ref{gauda}. We can define such a star-product e.g. by
\begin{eqnarray}
f(x)*''g(x)&:=\int d^2 z\;d^2 y\;\frac{1}{\pi^2 \det \theta} \exp[\frac{2i}{\theta}(x\wedge y+y\wedge z+z\wedge x)] \nonumber\\
&\Theta(l^2-(x-y)^2) \; \Theta(l^2-(x-z)^2) \; f(y) \; g(z),
\label{step1}
\end{eqnarray}
This definition is justified by that it respects the noncommutative symmetry
of space-time $O(1,1)\times SO(2)$. Analytical calculations are however very
restricted with the step-function cutoff and we cannot check for its associativity, let alone calculate
the commutator \eqref{comx} for the modification \eqref{step1}. Therefore we will be content with presenting the product \eqref{step1}
as a curiosity that might satisfy all of our requirements.

\section{Concluding remarks and discussion \label{conclrem}}

It has been shown here that the introduction of an extra exponential of derivatives into the differential
representation of the Weyl-Moyal star-product cannot remove the infinite nonlocality, inherent in
noncommutative theories using the Weyl-Moyal star-product.  This result is similar to the one found in
\cite{Galluccio:2009ss}, where it was shown that UV/IR mixing will remain, in theories with a more general star-product of the form
\be
f\star g=\frac1{(2\pi)^{\frac d2}}\int d^dp d^dq  e^{i p \cdot x} \tilde f(q)\tilde g(p-q) e^{\alpha(p,q)}. \label{intprodtran}
\ee
This product is required to remain associative, translationally invariant and satisfy the requirement
of the integral being a trace. That is,
\bea
\int  d^dx f \star  g&=&\int d^dx d^dp d^dq e^{\alpha(p,q)} e^{i p\cdot x} \tilde f(q) \tilde g (p-q)\nonumber\\
&=&\int d^dq e^{\alpha(0,q)} \tilde f(q) \tilde g (-q).
\eea
%Therefore it does seem that the infinite nonlocality of the commutator \eqref{comx} is indeed responsible
%for both the UV/IR mixing and the nonvanishing microcausality condition, although it remains to be proven
%in a strict sense.

Moreover, because the nonassociative Gaussian product \eqref{integral} does not remove the infinite nonlocality of noncommutative
quantum field theory, we thereby expect no analytical function in place of the Gaussian to do so. This leads one to propose that a possible way out could be placing a cutoff into the
integral representation of the Weyl-Moyal star-product. However, the very simple proposal \eqref{step1}, loses
much, if not all possibility of analytical calculation.

A different possibility out of the UV/IR problem could be provided by star-products on compact spaces, such as the
fuzzy sphere \cite{hopmad}.
%or other fuzzy spaces (for a review see e.g. \cite{Balbook}).
Since the standard cosmological model provides a
universe with a finite age and a finite size, it could be thought that space-time is compact.
% An implementation
%of compact space-time to 4-dimensional quantum field theory with some exactly defined symmetries has yet to be found, but it is a possibility not worth ruling out.
On the other hand, should
space-time be noncommutative and compact, the UV/IR problem could be solved but we would have no better understanding of the nonvanishing ETCR.
That is why the construction of a finite star-product in noncommutative field theories, if possible, would
enjoy a two-fold merit. It would provide a way out of both the UV/IR mixing problem and the nonvanishing ETCR. However, it should also be noted that were we able to modify
the Weyl-Moyal star-product to become nonlocal at a finite range, the connection of noncommutative
field theory to string theory would become vague. This follows because the Weyl-Moyal star-product with its infinite nonlocality is exactly the product found in the analysis of
\cite{SeiWitt}.

An additional remark we can make, regarding the construction of noncommutative field theories with a finite range of noncommutativity, concerns the Wightman-Vladimirov-Petrina (WVP) theorem \cite{WigVlaPet, BLT}.
A discussion of the implications of this theorem to noncommutative quantum field theory can also be found in \cite{Tureanu:2006ct}. The WVP-theorem is as follows:
If the commutator of observables in an ordinary quantum field theory satisfies a nonlocal causality condition of the form
\begin{eqnarray}
[\hat{O}(x), \hat{O}(y)] = 0, ~~~ \mathrm{for} ~~~ z_0^2-\sum_i \; z_i^2 < -l^2,~~~z=x-y,
\label{modcaus}
\end{eqnarray}
with $l$ a finite constant (fundamental) length,
then this implies the local microcausality condition
\begin{eqnarray}
[\hat{O}(x), \hat{O}(y)] = 0, ~~~ \mathrm{for} ~~~ z_0^2-\sum_i \; z_i^2 < 0,
\label{caus}
\end{eqnarray}
for all of space-time. At equal time, the relation \eqref{modcaus} is given by
\be
[\hat{O}(x), \hat{O}(y)]\big|_{x_0 = y_0} = 0, ~~~ \mathrm{for} ~~~ \sum_i \; z_i^2 > l^2,~~~z_i=x_i-y_i.
\label{mod2caus}
\ee

At a quick glance one might think that this theorem relates to our desire of constructing finite range noncommutativity, since the
Wightman functions  and axioms can be adapted to the noncommutative case \cite{AlvarezGaume:2003mb, Chaichian:2004qk}.
However, the finite noncommutativity we wish to obtain is of a different kind. We wish to have the two conditions:
\be
\mathrm{when}  \hspace{10pt} (\vec{x}_m - \vec{y}_m)^2 < l^2,   \hspace{10pt} \mathrm{to\ have} \hspace{10pt} [\hat{O}(x), \hat{O}(y)] = 0  \hspace{10pt} \mathrm{for}  \hspace{10pt} (x_0 - y_0)^2 - (x_1 - y_1)^2 \leq 0, \label{one}
\ee
and
\be
\mathrm{when} \hspace{10pt} (\vec{x}_m - \vec{y}_m)^2 > l^2,  \hspace{10pt} \mathrm{to\ have} \hspace{10pt} [\hat{O}(x), \hat{O}(y)] = 0  \hspace{10pt} \mathrm{for}  \hspace{10pt} (x - y)^2  < 0. \label{two}
\ee
Here, $\vec{x}_m = (x_2, x_3)$ and similarly for $\vec{y}_m$, i.e. the coordinates in the noncommutative plane.

Condition \eqref{one} is the condition for finite light-wedge causality, and gives us nonlocality within the range of $l$. Condition
\eqref{two} is the usual causality condition outside of the range $l$ of the finite noncommutativity. It can easily be seen that the two conditions \eqref{one} and \eqref{two}
cannot produce together the condition \eqref{modcaus} and thus the WVP-theorem is unrelated to the kind of finite noncommutativity we have been contemplating in this work.

%it is useful to recall the Wightman-Vladimirov-Petrina (WVP) theorem \cite{WigVlaPet, BLT}. Since the
%Wightman axioms and functions can be adapted to the noncommutative case \cite{AlvarezGaume:2003mb, Chaichian:2004qk}, it is interesting to explore the consequences of the
%WVP-theorem for theories with a finite range of noncommutativity.  These have already been briefly discussed in \cite{Tureanu:2006ct}.
%The WVP-theorem is as follows:

%At equal time, the relation \eqref{modcaus} is given by
%\be
%[\hat{O}(x), \hat{O}(y)]\big|_{x_0 = y_0} = 0, ~~~ \mathrm{for} ~~~ \sum_i \; z_i^2 > l^2,~~~z_i=x_i-y_i.
%\label{mod2caus}
%\ee
%This means that if we wish to construct a finite noncommutativity for which the commutator of observables
%vanishes for large enough space-like separation (i.e. outside of some specific range) as in \eqref{mod2caus}, this is equivalent to the observables commuting at all space-like separation, which
%is the commutative condition for microcausality. Therefore, one may expect that such a form of finite noncommutativity
%will lead to the loss of noncommutativity of the form \eqref{comx}, altogether. This is indeed the case.

%make it smarter!!!!
\noindent {\large \bf Acknowledgments}

We thank M. Chaichian, S. Saxell and A. Tureanu for illuminating discussions and for their help with the preparation of the manuscript.

%We also wish to thank Dr. S. Saxell for preparing a part of the manuscript.

\renewcommand{\baselinestretch}{1}\selectfont

\end{document}